\documentclass{PoS}

\usepackage{amssymb}
\usepackage{amsmath}
\usepackage{epsfig}
\usepackage{dcolumn}
\usepackage{rotating}
\usepackage{color}
\usepackage{comment}

\def\XXint#1#2#3{{\setbox0=\hbox{$#1{#2#3}{\int}$ }
\vcenter{\hbox{$#2#3$ }}\kern-.56\wd0}}

\newcommand*\xbar[1]{%
  \hbox{%
    \vbox{%
      \hrule height 0.5pt 
      \kern0.5ex
      \hbox{%
        \kern-0.1em
        \ensuremath{#1}%
        \kern-0.1em
      }%
    }%
  }%
}

\definecolor{rosso}{cmyk}{0,1,1,0.4}
\definecolor{rossos}{cmyk}{0,1,1,0.55}
\definecolor{rossoc}{cmyk}{0,1,1,0.2}
\definecolor{blu}{cmyk}{1,1,0,0.3}
\definecolor{blus}{cmyk}{1,1,0,0.6}
\definecolor{bluc}{cmyk}{1,1,0,0.1}
\definecolor{verde}{cmyk}{0.92,0,0.59,0.25}
\definecolor{verdec}{cmyk}{0.92,0,0.59,0.15}
\definecolor{verdes}{cmyk}{0.92,0,0.59,0.7}

\newcommand{\ba}{\begin{eqnarray}}
\newcommand{\ea}{\end{eqnarray}}
\newcommand{\be}{\begin{equation}}
\newcommand{\ee}{\end{equation}}
\newcommand{\bi}{\begin{itemize}}
\newcommand{\ei}{\end{itemize}}
\newcommand{\al}{\alpha}
\newcommand{\bt}{\beta}
\newcommand{\ga}{\gamma}

\newcommand{\la}{\lambda}

\newcommand{\Ga}{\Gamma}

\newcommand{\La}{\Lambda}

\newcommand{\cF}{{\cal F}}

\newcommand{\cK}{{\cal K}}

\newcommand{\cO}{{\cal O}}

\newcommand{\w}{\widetilde}

\newcommand{\st}{\stackrel}

\newcommand{\hp}{h^{\perp}}
\newcommand{\n}{\nabla}

\newcommand{\ra}{\rightarrow}
\newcommand{\Ra}{\Rightarrow}

\newcommand{\LF}{\left(}
\newcommand{\RF}{\right)}
\newcommand{\LT}{\left[}
\newcommand{\RT}{\right]}
\newcommand{\Ld}{\left.}
\newcommand{\Rd}{\right.}

\newcommand{\kb}{\bar{k}}
\newcommand{\pb}{\bar{p}}

\newcommand{\2}{\frac{1}{2}}

\newcommand{\4}{\frac{1}{4}}

\newcommand{\mt}{\mathtt}

\newcommand{\ie}{{\it i.e.\ }}

\newcommand{\non}{\nonumber\\}

\title{Towards understanding the ultraviolet behavior of quantum loops in  infinite-derivative theories of gravity}

\ShortTitle{Infinite-derivative theories of gravity}

\author{\speaker{Spyridon Talaganis}\\
        Lancaster University\\
        E-mail: \email{s.talaganis@lancaster.ac.uk}}

\abstract{In this proceedings (based on~\cite{Tal}), we will consider quantum aspects of a non-local, infinite-derivative scalar field theory - a {\it toy model} depiction of a covariant infinite-derivative, non-local extension of Einstein's general relativity which has previously been shown to be free from ghosts around the  Minkowski background.  The graviton propagator in this theory gets an exponential suppression making it {\it asymptotically free}, thus providing strong prospects of resolving various classical and quantum divergences. In particular, we will find that at $1$-loop, the $2$-point function is still divergent, but once this amplitude is renormalized by adding appropriate counter terms, the ultraviolet (UV) behavior of all other $1$-loop diagrams as well as the $2$-loop, $2$-point function  remains well under control.  We will go on to discuss how one may be able to generalize our computations and arguments to arbitrary loops.}

\FullConference{Proceedings of the Corfu Summer Institute 2014 "School and Workshops on Elementary Particle Physics and Gravity",\\
                3-21 September 2014\\
                Corfu, Greece }

\begin{document}

\section{Introduction}
\numberwithin{equation}{section}

Formulating a quantum theory of gravity~\cite{Veltman:1975vx,dewittQG,DeWitt:2007mi} remains one of the most outstanding challenges of high energy physics. While string theory (ST)~\cite{Polchinski:1998rr} remains the most popular candidate, other notable efforts include Loop Quantum Gravity (LQG)~\cite{Ashtekar,Nicolai:2005mc}, Causal Set approach~\cite{Henson:2006kf}, and ideas based on asymptotic safety~\cite{Weinberg:1980gg}. An interesting recurrent feature that appears in several of these approaches is non-locality. For instance, the entire formulation of LQG is based on non-local objects, such as Wilson loops and fluxes coming from the gravitational field.  Strings and branes of ST are, by their very definition, non-local objects. Even classically they do not interact with each other at a specific spatial point, but rather over a region in space. Not surprisingly, non-local structures are a common theme in stringy field theory (SFT) models. For instance, these appear in noncommutative geometry~\cite{noncom} \& SFT~\cite{Witten:1985cc},  for a review, see \cite{Siegel:1988yz}, and various {\it toy model}s of SFT such as $p$-adic strings~\cite{Freund:1987kt}, zeta strings~\cite{Dragovich:2007wb}, and strings quantized on a random lattice~\cite{Douglas:1989ve,Biswas:2004qu}. A key feature of these models is the presence of an {\it  infinite series of higher-derivative} terms incorporating the non-locality in the form of an {\it exponential kinetic} correction. Finally, it is also intriguing to note that similar infinite-derivative modifications have also been argued to arise in the asymptotic safety approach to quantum gravity~\cite{Krasnov}.


Accordingly, in \cite{Siegel:2003vt,Tseytlin:1995uq,Biswas:2005qr} attempts were made to construct ghost-free, infinite-derivative theories of gravity which may be able to resolve space-time singularities such as the ones present  inside the black holes and at the big bang. For instance, in~\cite{Biswas:2005qr} a non-singular bouncing
cosmological background was obtained within a class of infinite-derivative gravity theories, around which the sub and super-Hubble perturbations are well behaved and do not show
instabilities~\cite{Biswas:2010zk,Biswas:2012bp,Craps:2014wga}. In fact, such an action can also modify the famous Raychaudhuri's equation and alter the Hawking-Penrose singularity theorem~\cite{Conroy:2014dja}, which can yield a non-singular bouncing cosmology without violating the null energy conditions.

It was not until recently though, that concrete criteria for any covariant gravitational theory (including infinite-derivative theories) to be free from ghosts and tachyons around the Minkowski vacuum was obtained  by Biswas, Gerwick, Koivisto and Mazumdar (BGKM)~\cite{Biswas:2011ar,Biswas:2013kla}; see also~\cite{Modesto:2014eta} for a recent re-derivation of the same results using auxiliary field methods and~\cite{Eliezer:1989cr} for a detailed exposition of the problem of instabilities in infinite-derivative theories. In Ref.~\cite{Biswas:2011ar}, it was also shown how one can construct infinite-derivative theories of gravity where no new perturbative states are introduced and only the graviton propagator is modified by a multiplicative entire function. In particular, one can choose the entire function to correspond to the gaussian which suppresses the ultraviolet (UV) modes making the theory asymptotically free. For brevity we will refer to this case  as the BGKM model.

Given the prospects of the BGKM model at resolving the classical singularities of GR, see \cite{BT} for an overview, here we are going to explore the possibility of formulating a quantum theory of BGKM, and  the various challenges we need to overcome. For important works on slightly different approaches to quantizing gravity involving infinite-derivative interactions, see~\cite{Tomboulis,Modesto,BG,Anslemi,Moffat-qg}.

Let us start by  recalling the canonical examples of infinite-derivative actions, that appear in string literature. These can all be written as
\be
S=\int d^Dx \LT\2\phi \cK(\Box)\phi -V_{\mt{int}}(\phi)\RT\,,
\ee
where the kinetic operator $\cK(\Box)$ contains an infinite series of higher-derivative terms.
For instance, we find that $\cK(\Box)=-e^{\Box/M^2}$ for stringy {\it toy model}s based on $p$-adic numbers~\cite{Freund:1987kt}, or random lattices~\cite{Douglas:1989ve,Biswas:2004qu}, and $\cK(\Box)=-(\Box+m^2)e^{\Box/M^2}$  in SFT~\cite{Witten:1985cc}, where $m^2(<0)$ and $M^2(>0)$ are proportional to the string tension~\footnote{Here and hereafter, we are going to use  $(-+++)$ as our metric signature convention.}. Apart from its stringy origin, the above theories are interesting in their own right. Firstly, although these theories contain higher derivatives, they do not contain ghosts, at least perturbative. To see this explicitly, one can consider a fourth-order scalar theory with $\cK(\Box)=-\Box(1+{\Box\over m^2})$. The corresponding propagator reads
\be
\Pi(p^2)\sim \frac{1}{p^2(p^2 -m^2)}\sim \frac{1}{p^2 -m^2}-\frac{1}{p^2}\,.
\ee
From the pole structure of the propagator it is clear that the theory contains two physical states, but unfortunately the massive state has the ``wrong'' sign for the residue indicating that it is a ghost. Once interactions are included, it makes the classical theory unstable, and the quantum theory non-unitary (see~ Refs.~\cite{Efimov,Tomboulis:1983sw} regarding the issue of unitarity in infinite derivative theories). The stringy kinetic modifications combine to be an exponential, which is an {\it entire function} without any zeroes. In other words, it does not introduce any new states, ghosts or otherwise. Indeed, this property has been exploited to construct various non-local infinite-derivative theory and particle phenomenology models~\cite{Biswas:2014yia,Biswas:2009nx,Moffat:1990jj,Bluhm,Reddy}.

Secondly, as mentioned before, the infinite-derivative modification preserves a well known property of higher-derivative theories, that of making the quantum loop contributions better behaved in the UV. The stringy infinite-derivative scalar theories not only ameliorates the UV behavior, but the exponential suppression in the propagator actually makes all the quantum loops finite. Such calculations were used to provide evidence for several  stringy phenomena, such as Regge behavior~\cite{Biswas:2004qu} and thermal duality and Hagedorn transition~\cite{Biswas:2009nx}. It is then natural to wonder whether such non-local features can help in solving the quantum UV problem of gravity? In fact, Stelle, in Ref.~\cite{Stelle:1976gc} argued that the simplest higher-derivative theory of gravity, namely the fourth-order theory is already renormalizable, see also~\cite{Goroff:1985th,Goroff:1985sz}. Unfortunately,  the theory contains ghosts and is non-unitary. In contrast, the BGKM model provides  gravitational analogues of Eq.~\eqref{nlaction} where  the graviton propagator obtains an additional exponential suppression just as  the scalar models.

So, can this exponential infinite-derivative modification also solve the quantum UV problem of gravity by making all the Feynman loops finite? The answer is not  straightforward and our paper is essentially an {\it effort} to address this question. The main problem with the gravitational theories, as opposed to a scalar field theory, is that it is a gauge theory. And, one of the key features of gauge theories is that its free kinetic action is related to the interaction terms via the gauge symmetry. We will see that the exponential suppressions in propagators inevitably give rise to exponential enhancements in the vertex factors. Actually, this compensating interplay between propagators and vertices is not unique to infinite derivative theories, but any covariant theory of gravity, including Einstein's theory and Stelle's $4$th order gravity~\cite{Stelle:1976gc}, see also~\cite{Goroff:1985th,Goroff:1985sz}. In particular, the compensation between propagators and vertices is exact at the $1$-loop level making these contributions divergent as in GR. However, for higher loops, the superficial degree of divergence calculations is different from GR, because the counting is based on the pre-factors of the exponents rather than the degree of polynomial divergence; exponentials dominate any polynomial growth in the UV. In fact, a naive superficial divergence counting does suggest that diagrams with more than one loop  should be finite~\cite{Tomboulis,Modesto}.

The principal aim of this proceedings is to investigate the validity of such divergence counting in some details in a simplified {\it toy model} which retains the compensating feature of exponential suppression and enhancements between the propagator and interaction respectively. We will consider a  scalar field action, which maintains a combination of global scaling and  shift symmetries, similar to the residual
symmetry of gravity around the Minkowski background. Although, this symmetry manifests itself only at the level of  classical equations of motion, it will allow us to incorporate the compensating feature of exponential suppression and enhancement in propagators and interactions respectively, that is present in the full gravitational theory. We will consider a cubic interaction that respects the symmetry  and study $2$- and higher-point functions at $1$- and $2$-loops. We will first look at the vanishing external momentum limit, as they are technically easier to analyse and can already tell us whether a graph will be finite or not. It should be emphasized that we had to develop new techniques for regulating and evaluating the Feynman integrals. We will next look at the finite external momentum case, which is important in determining whether renormalizability arguments can be recursively pursued or not. Although, the cubic scalar interaction
inherits a bad IR behavior - being unbounded from below, still it serves as a very good example to study the UV aspects of the theory, which is the main focus of our paper. In particular, we will employ both hard cutoff and dimensional regularization techniques to regulate the loop integrals, and we will speculate how higher loops in these theories may also remain finite in the UV.

\section{Quantum gravity {\it toy model}}
\numberwithin{equation}{section}
\label{sec:toymodel}
\subsection{Superficial degree of divergence}\label{SDD}
Throughout this paper we will be interested in metric fluctuations, $h_ {\mu \nu}$, around the Minkowski background:
\be
g_{\mu \nu} = \eta_{\mu \nu} + h_{\mu \nu}\,.
\label{minkowski}
\ee
Gravity being a gauge theory only contains kinetic terms, \ie terms containing derivatives. In the case of GR all the terms contain two derivatives. In momentum space this means that the propagators behave as $k^{-2}$, while each vertex also comes with a $k^2$ factor. This is the compensating feature discussed in the introduction and is a hallmark of gauge theories.  Further in four dimensions, each momentum loop provides a $k^4$ factor in a quantum loop integral. The superficial degree of divergence of a Feynman diagram in GR is therefore given by (see~\cite{DeWitt:2007mi,Stelle:1976gc}):
\be
D = 4 L - 2 I + 2 V\,,
\ee
where $L$ is the number of loops, $V$ is the number of vertices, and $I$ is the number of internal propagators.
Using the topological relation:
\be
L = 1 + I - V\,,
\label{topological}
\ee
we get
\be
\label{eq:GR}
D = 2 L + 2\,.
\ee
Thus, the superficial degree of divergence increases as the number of loops increases, which is why GR is said to be non-renormalizable.

For Stelle's $4$th-order theory~\cite{Stelle:1976gc}, the graviton propagator goes as $\sim k^{-4}$, while the vertices $\sim k^4$, leading to a constant degree of divergence formula
\begin{equation}
D = 4\,.
\end{equation}
In other words, the degree of divergence does not increase with loops which enabled Stelle to prove that such a theory is renormalizable. Unfortunately, such a theory also contains a Weyl ghost which makes the theory non-unitary. As explained before, we will follow a different approach where we will introduce an infinite series of higher-derivative operators in a way that doesn't introduce any new states, ghosts or otherwise. We will see that the divergence counting will also be different as it will be based on the exponents rather than the degree of the polynomial momentum dependences.

\subsection{Infinite-derivative gravitational action}\label{intro-1}
The ``simplest'' infinite-derivative action that can modify the propagator of the graviton  without introducing any new states  is of the form~\cite{Biswas:2011ar,Biswas:2013kla}
\be
\label{action}
S = S_{EH} + S_{Q}\,,
\ee
where $S_{EH}$ is the Einstein-Hilbert action,
\be
\label{eq:EH}
\int \mathrm{d} ^ 4 x \, \sqrt{-g} \, \frac{R}{2}\,,
\ee
and $S_{Q}$ is given by~\footnote{Around Minkowski space or in any maximally symmetric background it can be shown that ${\cal F}_3$ is redundant,
see~\cite{Biswas:2011ar,Biswas:2013kla}.}:
\be
\label{nlaction}
S_{Q} = \int \mathrm{d} ^ 4 x \, \sqrt{-g} \LT R \cF _ {1} (\Box) R + R _ {\mu \nu} \cF _ {2} (\Box) R ^ {\mu \nu} + R _ {\mu \nu \lambda \sigma} \cF _ {3} (\Box) R ^ {\mu \nu \lambda \sigma}\RT\,,
\ee
where the $\cF_i$'s are analytic functions of $\Box$ (the covariant d'Alembertian operator):
\be
\cF_{i} (\Box) = \sum _ {n=0}^{\infty} f_{i_n} \Box^{n}\,,
\label{quadratic}
\ee
satisfying~\footnote{For other forms of infinite-derivative gravity theories which contain an additional scalar degree of freedom, see~\cite{Biswas:2013kla,BT}.}
\be
\label{eq:kuku}
2\cF_1+\cF_2+2\cF_3=0\,,
\ee
and the constraint that the combination
\be
\label{eq:a}
a (\Box) = 1 - \frac{1}{2} \mathcal{F} _ {2} (\Box) \Box - 2 \mathcal{F} _ {3} (\Box) \Box\,,
\ee
is an {\it entire function}, with no zeroes. In Eq.~\eqref{quadratic}) $f_{i_n}$'s  are real coefficients. Eqs.~\eqref{action}-\eqref{eq:a} defines the BGKM gravity models. The classical equations of motion have been
studied for the above action~\cite{Biswas:2013cha}, and shown to be free from black-hole type of singularities for ``small'' central masses. In this proceedings, therefore, we take the next logical step of investigating the quantum UV behavior of these theories.

For BGKM type models, the quadratic (in $h_{\mu\nu}$) or ``free'' part of the action simplifies considerably, and one obtains:
\be
S_{\mt{free}}={M_p^2\over 2}\int d^4x\ h^{\perp\mu\nu}\Box a(\Box)\hp_{\mu\nu}\,,
\ee
where $\hp_{\mu\nu}$
is the transverse traceless spin $2$ graviton mode, satisfying:
\be
\n^{\mu}\hp_{\mu\nu}=g^{\mu\nu}\hp_{\mu\nu}=0\,.
\ee
This leads to the propagator (in Feynman gauge)~\cite{Biswas:2011ar,Biswas:2013kla}
\be
\Pi(k^2) = - \frac{i}{k^2a(-k^2)}\LF {\cal P}^2 - \frac{1}{2} {\cal P}_s ^0  \RF=-\frac{i}{a(-k^2)} \Pi_{GR}\,,
\ee
for the physical degrees of freedom for the graviton (see~\cite{Biswas:2013kla,peter} for the definitions of the spin projector operators ${\cal P}^2$ and ${\cal P}_s ^0$).

Ideally, we should now compute the interaction terms for our non-local gravity theory and then use it to compute the Feynman diagrams. This however turns out to be an extremely challenging task for several reasons: Gravitational theories are all order theories and therefore contain interactions of all orders in $ h_{\mu\nu}$, and computing all these interactions is well beyond the scope of the current paper. While one can argue that all the terms which are higher order in fields have additional Planck suppressions, and that therefore the most relevant piece in the low energy approximation comes from the cubic terms, unfortunately even computing the complete cubic interactions for an action such as Eq.~\eqref{nlaction} is challenging. Moreover, the expressions are rather complicated making further progress in evaluating Feynman diagrams very difficult.

\subsection{Motivating scalar {\it toy model} of quantum gravity from symmetries}
\label{sec:scalartoymodel}
It is well known that the field equations of GR exhibit a global scaling symmetry,
\be
g_{\mu\nu}\ra \la g_{\mu\nu}\,.
\label{scaling}
\ee
When we expand the metric around the
Minkowski vacuum, Eq.(\ref{minkowski}), the scaling symmetry translates to a symmetry for $h_{\mu\nu}$, whose infinitesimal version is given by
\be
h_{\mu\nu} \to (1 + \epsilon) h_{\mu\nu} + \epsilon\eta_{\mu\nu}\,.
\ee
While we do not expect the scaling symmetry to be an unbroken fundamental symmetry of nature, the symmetry serves a rather useful purpose for us.
It relates the free and interaction terms just like gauge symmetry does. Thus, we are going to use this combination of shift and scaling symmetry:
\be
\phi \to (1 + \epsilon) \phi + \epsilon\,,
\label{scale-shift}
\ee
to arrive at a scalar {\it toy model} whose propagator and vertices preserve the compensating nature found in the full BGKM gravity.
Inspired by the discussion in the previous section, we will now consider a scalar {\it toy model} with a string field theory type free action:
\be
\label{free}
S_{\mt{free}} = \frac{1}{2}\int \mathrm{d}^4 x \, \LF  \phi \Box a(\Box) \phi\RF\,,
\ee
where for the purpose of this paper, we are going to choose~\cite{Biswas:2011ar,Biswas:2013kla}:
\be
a (\Box) = e^{- \Box / M ^ {2}}\,,
\ee
where $M$ being the mass scale at which the non-local modifications become important. In general, one is free to choose any entire function,
while keeping in mind that $a(k^2)\rightarrow 1$ for the IR momentum, $k\rightarrow 0$, in order to recover the propagator of the usual GR.  Note that the
sign of $a(\Box)$ is also crucial in order to recover the correct Newtonian potential as shown in Ref.~\cite{Biswas:2011ar,Biswas:2013kla}.

The symmetry, Eq.~(\ref{scale-shift}), then uniquely fixes the cubic interaction term,
\be
\label{int}
S_{\mt{int}} = \frac{1}{M_p} \int \mathrm{d} ^ 4 x \, \LF \4 \phi \partial _ {\mu} \phi \partial ^ {\mu} \phi + \4 \phi \Box \phi a(\Box) \phi - \4 \phi \partial _ {\mu} \phi a(\Box)  \partial ^ {\mu} \phi \RF\,,
\ee
up to integrations by parts. Our  {\it toy model} action is then be given by:
\be
\label{eq:action}
S_{\mt{scalar}} = S _ {\mt{free}} + S _ {\mt{int}}\,.
\ee
It is now time to revisit the superficial degree of divergence for this {\it toy model}.

Since an exponential suppression always dominates over a polynomial growth, the naive expectation is that as long as the exponentials come with a negative power, the integrals should converge. Thus, rather than computing the power of polynomial divergence in momentum, we are really interested in calculating the pre-factor in the exponent, and this radically changes the counting of the superficial degree of divergence.

Since every propagator comes with an exponential suppression, see Eq.~(\ref{free}), while every vertex comes with an exponential enhancement, Eq.~(\ref{int}), the superficial degree of divergence counting in exponents is given by
\be
E = V - I\,.
\ee
By using the topological relation, Eq.~\eqref{topological}, we obtain:
\be
\label{eq:E}
E = 1 - L\,.
\ee
Thus, except for the $L = 1$ loop, $E < 0$, and the corresponding loop amplitudes are superficially convergent.

\section{Divergence structure}\label{sec:zero-momentum}
\numberwithin{equation}{section}


We will study the UV properties of a toy scalar field theory. The scalar theory preserves a combination of the conformal and  shift symmetries (like gravity around Minkowski space-time),  $\phi \ra (1+\epsilon)\phi + \epsilon$. This ensures that the most relevant property of gravitational theories when it comes to studying quantum UV properties, \ie the compensatory nature of the exponential suppressions and enhancements in the propagators and  vertex factors respectively, is retained.
The action containing cubic interactions is given by
\be
\label{eq:action1}
S = \frac{1}{2}\int \mathrm{d}^4 x \, \LF  \phi \Box a(\Box) \phi\RF + \frac{1}{M_p} \int \mathrm{d} ^ 4 x \, \LF \4 \phi \partial _ {\mu} \phi \partial ^ {\mu} \phi + \4 \phi \Box \phi a(\Box) \phi - \4 \phi \partial _ {\mu} \phi a(\Box)  \partial ^ {\mu} \phi \RF\ .
\ee

The propagator in momentum space is given by
\be
\Pi (k ^ 2)= \frac{- i}{k^2 e ^ {\kb ^ 2}},
\ee
 while the vertex factor $V(k_{1},k_{2},k_{3})$ is given by
\be
\label{eq:V}
V (k _ {1}, k _ {2}, k _ {3}) = \frac{i}{4} \LF k_{1}^2+k_{2}^2+k_3^2 \RF \LT 1 -  e ^ {\kb _ {1} ^ {2}} -  e ^ {\kb _ {2} ^ {2}} - e ^ {\kb _ {3} ^ {2}}\RT \Ra V(k)\equiv V (k, -k, 0) = - i k ^ {2} e ^ {\kb ^ {2}}\ .
\ee
The barred  momenta denote that the momenta have been divided by the mass scale $M$.
The $1$-loop, $N$-point function with zero external momenta evaluates to
\be
\label{eq:N}
\Ga_N = (-1) ^ {N} \frac{i \Lambda ^4}{32 M _ {p} ^ {N} \pi ^ 2}\ ,
\ee
where $\La$ is a hard cutoff. The power of divergence doesn't change with $N$ and is exactly as one would expect from the counting argument (\ref{eq:E}).

If we evaluate the two-loop Feynman diagrams for zero external momenta, we again find a $\La^4$ divergence as would be expected since the 2-loop diagrams contain $1$-loop subdivergences. Again this suggests  that we do not get any additional divergences as we proceed from $1$-loop to $2$-loops, corroborating (\ref{eq:E}).  However, this does not guarantee renormalizability. To prove renormalizability we must make sure that after we eliminate the 1-loop divergences (by possibly adding suitable counterterms), the remaining external momentum dependence of the 1-loop diagrams is, at least, milder than the bare vertex in the UV to make the higher loops convergent. With this in mind, let us check the external momentum dependence of the $1$-loop, $2$-point function:
\be
\Ga_{2,1}(p^2) = \frac{i}{2 i ^ {2} M _ {p} ^ {2}} \int \frac{\mathrm{d} ^ 4 k}{(2 \pi) ^ {4}} \, \frac{V ^ {2} (-p, \frac{p}{2} + k, \frac{p}{2} - k)}{(\frac{p}{2} + k) ^ {2} (\frac{p}{2} - k) ^ {2} e ^ {\LF\frac{\pb}{2} + \kb\RF ^ {2}} e ^ {\LF\frac{\pb}{2} - \kb\RF ^ {2}}}\,\ .
\label{2pt-ext}
\ee

The $1$-loop, $2$-point contribution schematically reads:
\be
\Ga_{2,1}(p^2)+\Ga_{2,1,ct}(p^2)=\Ga_{2,1\mt{r}}(p^2)= {i M^4\over M_p^2}f(\pb^2)\,,
\ee
where
\begin{align}
f (\pb) & = \frac{\pb ^ {4}}{128 \pi ^ {2}} \left(- \log \left(\frac{\pb^2}{4 \pi}\right) - \gamma +2 \right) \non
& + \frac{e^{-\pb^2}}{512 \pi ^2 \pb^2} \LT \vphantom{\text{Ei}\LF \frac{\pb^2}{2}\RF}-2 e^{\pb ^ 2} \LF e ^ {2 \pb ^2} - 1 \RF \pb^6 Ei\LF-\pb^2 \RF + \LF e ^ {\pb^2} - 1\RF \LF \vphantom{\text{Ei}\LF \frac{\pb^2}{2}\RF}-2 \LF \pb^4 + 3 \pb^2 + 2 \RF \Rd \Rd \non
& + \Ld \Ld  \LF e^{\frac{3 \pb^2}{2}} -e^{\frac{\pb^2}{2}} \RF \LF 2\pb^4 + 5 \pb^2 + 4 \RF + e ^{\pb^2} \LF e^{\pb^2} -1 \RF \pb^6 Ei \LF - \frac{\pb^2}{2} \RF + 2 e^{\pb^2} \LF 7 \LF \pb^4 + \pb^2 \RF +2 \RF \RF \RT \,.
\end{align}
$f(\pb)$ is a regular analytic function of $\pb$ which grows as $e^{3\pb^2/2}$ as $\pb \ra \infty$.
In dimensional regularization scheme we obtain an $\epsilon ^ {-1}$ pole, as expected, which can be eliminated using suitable counterterms. The problem however is that as $p\ra \infty$, $\Ga_{2,1}(p^2) $ now diverges exponentially as $e^{\frac{3\pb^2}{2}}$! This, for instance, would make the $2$-loop diagram, which has a renormalized $1$-loop 2-pt subdiagram, divergent. So, is all hope of renormalizability lost?

It is well known that a $1$-loop, $2$-point insertion in any graph can be replaced by a sequence of such insertions separated by the bare propagator. The sequence of all the graphs gives rise to a geometric series which can be summed in the appropriate regime and then analytically continued to the entire complex $p^2$-plane giving rise to the ``dressed'' propagator,
\be
\w{\Pi}(p^2) = \frac{\Pi(p^2)}{1-\Pi(p^2)\Ga_{2,r}(p^2)} \ .
\ee
Remarkably, now the fact that at large momentum, $\Ga_{2,r}$ grows even more strongly than $e^{\pb^2}\sim \Pi^{-1}$, makes the dressed propagators  more exponentially suppressed than their bare counterparts: $ \w{\Pi}(p^2)\ra  e^{-\frac{3\bar{p}^2}{2}} $.
In particular, this makes all $1$-loop graphs, higher than 2-point,  UV-finite! The UV part of the $2$-loop integrals become finite too once the 1-loop 2-pt contribution is renormalized.

\section{Higher vertices and prospects for a finite theory}
We have just now seen how strong exponential suppression of the dressed propagator can make the $1$-loop and $2$-loop integrals finite. We believe that most likely this remarkable feature continues to higher loops. The basic reason is - even for the $1$-loop diagrams, the suppression coming from the propagators is stronger than the enhancements coming from the vertices. This ensures two things - first it makes the loops finite, and second the UV growth of the finite diagrams with respect to the external momenta becomes weaker in every subsequent loops. Thus, finiteness of higher loops is  guaranteed recursively. A rigorous proof of the above statement is well beyond the scope of the present paper, but we will now sketch heuristic arguments to demonstrate finiteness of the  particular set of $2$- and $3$-point diagrams that can be constructed out of lower-loop $2$- and $3$-point diagrams.


The basic approach is the following - in order to understand whether any diagram converges in the UV or not, we only need to keep track of the exponential momentum dependences. We already know that the dressed propagators, represented by the shaded blobs, decay in the UV as $e^{-{3\kb^2}/{2}}$. Conservatively, we are therefore going to assume
$\Pi(k^2)\st{UV}{\longrightarrow} e^{-{3\kb^2}/{2}}$. The $3$-point function (represented by the dark blobs) can, on the other hand, be written as
\be
\label{eq:external}
\Ga_3\st{UV}{\longrightarrow}\sum_{\al,\bt,\ga} e^{\al\pb_{1}^{2}+\bt\pb_{2}^{2}+\ga\pb_{3}^{2}}\ \,,
\ee
with  the convention
\be
\al\geq\bt\geq\ga \,,
\ee
where $p_1,~p_2,~p_3$ are the three external momenta.  This is because once all the (lower-) loop sub diagrams have been integrated out, what remains are expressions in terms of the corresponding external momenta. Some of these external momenta can then become the internal loop momentum in a subsequent higher loop diagram.

The sum over the exponents $\{\al,~\bt,~\ga\}$ in Eq. (\ref{eq:external}) indicates that there could be many different exponential terms including the permutations needed to symmetrize the vertices over the three internal momenta. We are going to assume that these exponents satisfy certain properties, up to say $(n-1)$-loops. These conditions will allow us to demonstrate that the loops remain finite. Moreover, we will  recursively argue that these properties are also satisfied in
the $n$-th loop.
\subsubsection{$2$-point diagram}
\label{sec:2-point}
First, let us look at the zero external momentum limit.
It is easy to see that the most divergent UV part of the $2$-point diagram reads
\be
\label{eq:514}
\Ga_{2,n}{\longrightarrow}\int \frac{\mathrm{d} ^ 4 k}{(2 \pi) ^ 4} {e^{(\al_1+\al_2+\bt_1+\bt_2)\kb^{2}}\over e^{3\kb^2}}\,,
\ee
where $k$ is the loop momentum variable. We've got two propagators $e^{\frac{3 \kb^2}{2}}$   while the (most divergent UV parts of the) vertex factors originating from lower-loop diagrams are $e ^ {\al_{1} \kb^2 + \bt_{1}\kb^2}$ and $e ^ {\al_{2} \kb^2 + \bt_{2}\kb^2}$ (we get no $\ga_{1}$, $\ga_{2}$ terms in the exponents, since the external momenta are set equal to zero). Clearly, the integral is finite as long as
\be
\al_{i}+\bt_{i}<\frac{3}{2}\,,
\label{condition1}
\ee
where $i=1, 2$.  One can check that the same condition ensures finiteness of the diagram even when one includes non-zero external momenta.
\subsubsection{$3$-point diagram}
First, let us check whether the $3$-point diagram is finite or not for zero external momenta. Again the most divergent UV contribution comes when the momentum associated with exponents, $\al$'s and $\bt$'s, run in the internal loop giving rise to
\be
\label{eq:516}
\Ga_{3,n}{\longrightarrow}\int \frac{\mathrm{d} ^ 4 k}{(2 \pi) ^ 4} {e^{(\al_1+\al_2+\al_3+\bt_1+\bt_2+\bt_3)\kb^2}\over e^{\frac{9\kb^2}{2}}}\,,
\ee
where $k$ is the loop momentum variable. Similarly to the argument for the 2-pt function, we've got three propagators $e^{\frac{3 \kb^2}{2}}$,   while the (most divergent UV parts of the) vertex factors originating from lower-loop diagrams are $e ^ {\al_{1} \kb^2 + \bt_{1}\kb^2}$, $e ^ {\al_{2} \kb^2 + \bt_{2}\kb^2}$ and $e ^ {\al_{3} \kb^2 + \bt_{3}\kb^2}$.  Again the integral converges as long as Eq. (\ref{condition1}) is valid.

To prove the validity of Eq. (\ref{condition1}), let us try to find out how one can get the largest exponents for the external momenta. First, let us consider how one can get the largest sum of all the exponents, {\it i.e.}, $\al+\bt+\ga$. Although, all the arguments below can be conducted for three different sets of exponents in the three $3$-point vertices making up the $1$-loop triangle, for simplicity, here we will look at what happens when all the three vertices have the same exponents. Clearly, the best way to obtain the largest exponents for the external momenta is to have the $\al$ exponent correspond to the external momenta. For a symmetric distribution of $(\bt, \ga)$ among the internal loops, we get
\be
\label{eq:crucial}
\Ga_{3,n}{\longrightarrow}\int \frac{\mathrm{d} ^ 4 k}{(2 \pi) ^ 4} {e^{\al^{n-1}(\pb_1^2+\pb_2^2+\pb_3^2)}\over e^{[{3\over 2}-\bt^{n-1}-\ga^{n-1}][3\kb^2+\frac{1}{3}(\pb_1^2+\pb_2^2+\pb_3^2)]} }\,,
\ee
where $p_{1}$, $p_{2}$, $p_{3}$ are the external momenta for the $1$-loop triangle, and  the superscript in the $\al,\bt,\ga$ indicates that these are coefficients that one obtains from contributions up to $n-1$  loop level.  Before proceeding to obtain the $n$th loop coefficients, let us briefly explain how we got Eq.~\eqref{eq:crucial}. Assuming symmetrical routing of momenta in the $1$-loop triangle, we get the propagators $e^{-\frac{3}{2} \LF \kb + \frac{\pb_{1}}{3} - \frac{\pb _ {2}}{3} \RF^2}$, $e^{-\frac{3}{2} \LF \kb + \frac{\pb_{2}}{3} - \frac{\pb _ {3}}{3} \RF^2}$ and $e^{-\frac{3}{2} \LF \kb + \frac{\pb_{3}}{3} - \frac{\pb _ {1}}{3} \RF^2}$, and the vertex factors $e ^ {\al^{n-1} \pb _ {1} ^ {2} + \bt^{n-1} \LF \kb + \frac{\pb _ {3}}{3} - \frac{\pb _ {1}}{3} \RF^2 + \ga^{n-1} \LF \kb + \frac{\pb _ {1}}{3} - \frac{\pb _ {2}}{3} \RF ^2}$, $e ^ {\al^{n-1} \pb _ {2} ^ {2} + \bt^{n-1} \LF \kb + \frac{\pb _ {1}}{3} - \frac{\pb _ {2}}{3} \RF^2 + \ga^{n-1} \LF \kb + \frac{\pb _ {2}}{3} - \frac{\pb _ {3}}{3} \RF ^2}$ and $e ^ {\al^{n-1} \pb _ {3} ^ {2} + \bt^{n-1} \LF \kb + \frac{\pb _ {2}}{3} - \frac{\pb _ {3}}{3} \RF^2 + \ga^{n-1} \LF \kb + \frac{\pb _ {3}}{3} - \frac{\pb _ {1}}{3} \RF ^2}$. Conservation of momenta then yields Eq. (\ref{eq:crucial}).

By integrating Eq. \eqref{eq:crucial}, we have
\be
\al^n=\bt^n=\ga^n=\al^{n-1}+\frac{1}{3}(\bt^{n-1}+\ga^{n-1})-\2 \,.
\label{alpha-n}
\ee
In particular, for the $1$-loop, $3$-point graph, one has to use the $3$-point bare vertices: $\al^0=1$ and $\bt^0=\ga^0=0$. One then obtains
\be
\al^1=\bt^1=\ga^1=\2 \,,
\ee
leading to an overall symmetric vertex: $e^{\2(\pb_1^2+\pb_2^2+\pb_3^2)}$ and $\al ^ {1}+\bt^{1}+\ga^{1} = \frac{3}{2}$. Since we expect the exponents to decrease as we increase loops, we therefore conjecture that the sum of exponents satisfies the inequality,
\be
\al^{n} +\bt^{n} +\ga^{n} \leq \frac{3}{2} \,.
\label{condition2}
\ee
From Eq.~\eqref{alpha-n}, we see that this is satisfied provided a further condition is satisfied by the exponents, \ie:
\be
\al^{n-1}+\frac{1}{3}(\bt^{n-1}+\ga^{n-1})\leq 1 \,.
\label{condition3}
\ee
To summarize, so far we have shown that if, up to $n-1$ loops, inequality Eq.~\eqref{condition3}
is satisfied, then, at the $n$-th loop, Eq.~\eqref{condition2} is also satisfied. To complete the recursive proof, we must argue that Eq.~\eqref{condition3} is also satisfied at $n$-loops. For the loop contribution we are discussing, we have
\be
\al^{n}+\frac{1}{3}(\bt^{n}+\ga^{n})=\frac{5}{3}\LT\al^{n-1}+\frac{1}{3}(\bt^{n-1}+\ga^{n-1})-\2\RT\leq \frac{5}{6}<1\,,
\ee
and Eq.~\eqref{condition3}  is indeed satisfied.

One may wonder whether there are other ways of distributing the exponents which could violate Eq.~\eqref{condition2}. For instance, one can try to maximize $\al^n$ by distributing
$\al^{n-1}$ in two of the vertices to run along the internal loop. However, one can check that Eq.~\eqref{condition3} still remains valid.

The final point is that the sum of the exponents is maximized by distributing the largest exponents to all the external momentum, in particular making all the exponents positive thereby ensuring that Eq.~\eqref{condition1} follows from Eq.~\eqref{condition2}. While we do not yet have a rigorous proof of these above arguments, in all the cases we have looked at so far, the inequalities, Eq.~\eqref{condition1}, Eq.~\eqref{condition2} and Eq.~\eqref{condition3} seem to hold up.

\section{Summary \& future research directions}\label{sec:conclusions}
\numberwithin{equation}{section}

In this proceedings, we studied the quantum loops for an infinite-derivative scalar field theory action as a toy model to mimic the UV properties of the BGKM
gravity~\cite{Biswas:2011ar}. Expanding the BGKM action around the Minkowski vacuum one can obtain, for instance,  the ``free'' part that determines the propagator from the $\cO (h^2)$ terms,  while the $\cO (h^3)$ terms determines the cubic interaction vertices. Unfortunately, $\cO (h^3)$  terms  are technically challenging and some of the expression involves double sums. Instead of getting involved with too many technicalities, we therefore chose to work with a simple toy model
action, Eq.~(\ref{eq:action}), that respects a combination of the shift and scaling symmetry at the level of equation of motion that lets us capture some of  the essential features  of BGKM gravity such as the compensating nature of the exponential suppression in the propagator and the exponential enhancement in the  vertex factor.

We derived the Feynman rules for our toy model action, {\it i.e.}, the propagator and the vertex factors. Consequently, we computed the $1$-loop, $2$-point diagram, both with zero and arbitrary external momenta, which gives a $\La^4$ divergence, where $\La$ is a momentum cutoff. The $2$-loop diagrams with zero external momenta also give a $\La^4$ divergence, suggesting that we do not get new divergences as we proceed from $1$-loop to $2$-loop. We repeated our
$1$-loop and $2$-loop computations with external momenta, and we paid extra care in understanding the  $1$-loop, $2$-point function which appears as a subdivergence in higher-loop diagrams. Typically, in the $1$-loop, $2$-point function, we obtain a $e^{\frac{3 \pb^2}{2}}$ external momentum dependence in the UV
which indicates that for $\pb \to \infty$, the $1$-loop, $2$-point function tends to infinity. This may appear as an initial setback, but actually, this external momentum dependence is what , we believe, makes all higher-loop and higher-point diagrams finite, once the bare propagators are replaced with the dressed propagators.

This becomes possible because  the exponential suppression in the dressed propagator overcomes the exponential enhancement originating from the vertices. The $1$-loop, $N$-point functions with zero external momenta become UV-finite, as do the $2$-loop integrals for vanishing external momenta. We believe that, even in the case of arbitrary external momenta, our results will not change; the higher-loop diagrams also become UV-finite with the use of the dressed propagators. The basic reason is that, even for the $1$-loop diagrams, the suppression coming from the propagators is stronger than the enhancements coming from the vertices. This ensures two things - first it makes the loops finite, and second the UV growth of the finite diagrams with respect to the external momenta becomes weaker in every subsequent loops. Thus, finiteness of higher loops is  guaranteed recursively.

To illustrate this general argument, we considered the finiteness of $n$-loop, $2$-point and $3$-point diagrams that can be constructed out of lower-loop $2$-point and $3$-point diagrams. This is already an
encouraging sign for an infinite-derivative action of scalar toy model, which can now make higher loops finite, giving us a ray of hope to tackle the problem in full glory for the BGKM gravity. However,
as a future computation, it would be interesting to first demonstrate that the finiteness of the diagrams hold to all orders in loops for any $N$-point diagrams. A full proof even for an infinite-derivative  toy model is beyond our scope, and we will carry on this computation elsewhere.

\section{Acknowledgments}

ST is supported by a scholarship from the Onassis Foundation.


\begin{thebibliography}{99}

\bibitem{Tal}
 S.~Talaganis, T.~Biswas and A.~Mazumdar,
  ``{\it Towards understanding the ultraviolet behavior of quantum loops in infinite-derivative theories of gravity,}''
  arXiv:1412.3467 [hep-th].
  
\bibitem{Veltman:1975vx}
  M.~J.~G.~Veltman,
  ``{\it Quantum Theory of Gravitation,}''
  {\it Conf.\ Proc.\ C} {\bf 7507281}, 265 (1975).

\bibitem{dewittQG}
 B.~S.~DeWitt,
  ``{\it Quantum Theory of Gravity. 1. The Canonical Theory,}''
  {\it Phys.\ Rev.} {\bf 160}, 1113 (1967).\\
   B.~S.~DeWitt,
  ``{\it Quantum Theory of Gravity. 2. The Manifestly Covariant Theory,}''
  {\it Phys.\ Rev.} {\bf 162}, 1195 (1967).\\
   B.~S.~DeWitt,
 ``{\it Quantum Theory of Gravity. 3. Applications of the Covariant Theory,}''
 {\it Phys.\ Rev.} {\bf 162}, 1239 (1967).

\bibitem{DeWitt:2007mi}
  B.~S.~DeWitt and G.~Esposito,
  ``{\it An Introduction to quantum gravity,}''
  {\it Int.\ J.\ Geom.\ Meth.\ Mod.\ Phys.} {\bf 5}, 101 (2008)
  [arXiv:0711.2445 [hep-th]].


\bibitem{Polchinski:1998rr}
  J.~Polchinski,
  ``{\it String theory. Vol. 2: Superstring theory and beyond,}''
  Cambridge, UK: Univ. Pr. (1998) 531p.

\bibitem{Ashtekar}
A.~Ashtekar,
  ``{\it Introduction to loop quantum gravity and cosmology,}''
  {\it Lect.\ Notes Phys.} {\bf 863}, 31 (2013).

\bibitem{Nicolai:2005mc}
for a review, see:  H.~Nicolai, K.~Peeters and M.~Zamaklar,
  ``{\it Loop quantum gravity: An Outside view,}''
  {\it Class.\ Quant.\ Grav.}  {\bf 22}, R193 (2005)
  [hep-th/0501114].


  \bibitem{Henson:2006kf}
for a review, see:  J.~Henson,
  ``{\it The Causal set approach to quantum gravity,}''
  In *Oriti, D. (ed.): Approaches to quantum gravity* 393-413
  [gr-qc/0601121].


\bibitem{Weinberg:1980gg}
  S.~Weinberg,
  ``{\it Ultraviolet Divergences In Quantum Theories Of Gravitation,}''
  Hawking, S.W. (Cambridge Univ. (UK)); Israel, W. (Alberta Univ., Edmonton (Canada). Theoretical Physics Inst.) (eds.); p. 790-831; ISBN 0 521 22285 0; 1979; p. 790-831; University Press; Cambridge

\bibitem{noncom}
A.~Smailagic and E.~Spallucci,
  ``Lorentz invariance, unitarity in UV-finite of QFT on noncommutative
spacetime,''
  {\it J.\ Phys.\ A} {\bf 37}, 1 (2004)
  [{\it Erratum-ibid.\ A} {\bf 37}, 7169 (2004)]
  [hep-th/0406174].\\
E.~Spallucci, A.~Smailagic and P.~Nicolini,
  ``{\it Trace Anomaly in Quantum Spacetime Manifold,}''
  {\it Phys.\ Rev.\ D} {\bf 73}, 084004 (2006)
  [hep-th/0604094].\\
   P.~Nicolini, A.~Smailagic and E.~Spallucci,
  ``{\it Noncommutative geometry inspired Schwarzschild black hole,}''
  {\it Phys.\ Lett.\ B} {\bf 632}, 547 (2006)
  [gr-qc/0510112].\\
  P.~Nicolini,
  ``{\it Noncommutative Black Holes, The Final Appeal To Quantum Gravity: A Review,}''
  {\it Int.\ J.\ Mod.\ Phys.\ A} {\bf 24}, 1229 (2009)
  [arXiv:0807.1939 [hep-th]].

  \bibitem{Witten:1985cc}
  E.~Witten,
  ``{\it Noncommutative Geometry and String Field Theory,}''
 {\it Nucl.\ Phys.\ B} {\bf 268}, 253 (1986).


\bibitem{Siegel:1988yz}
W.~Siegel,
  ``{\it Introduction to string field theory,}''
  hep-th/0107094.

  \bibitem{Freund:1987kt}
  P.~G.~O.~Freund and M.~Olson,
  ``{\it Nonarchimedean Strings,}''
 {\it Phys.\ Lett.\ B} {\bf 199}, 186 (1987).\\
  P.~G.~O.~Freund and E.~Witten,
  ``{\it Adelic String Amplitudes,}''
 {\it Phys.\ Lett.\ B} {\bf 199}, 191 (1987).\\
  L.~Brekke, P.~G.~O.~Freund, M.~Olson and E.~Witten,
  ``{\it Nonarchimedean String Dynamics,}''
 {\it Nucl.\ Phys.\ B} {\bf 302}, 365 (1988).\\
  P.~H.~Frampton and Y.~Okada,
  ``{\it Effective Scalar Field Theory of $P^-$adic String,}''
 {\it Phys.\ Rev.\ D} {\bf 37}, 3077 (1988).



  \bibitem{Dragovich:2007wb}
  B.~Dragovich,
  ``{\it Zeta strings,}''
  hep-th/0703008.




  \bibitem{Douglas:1989ve}
  M.~R.~Douglas and S.~H.~Shenker,
  ``{\it Strings in Less Than One-Dimension,}''
 {\it Nucl.\ Phys.\ B} {\bf 335}, 635 (1990).\\
  D.~J.~Gross and A.~A.~Migdal,
  ``{\it Nonperturbative Solution of the Ising Model on a Random Surface,}''
 {\it Phys.\ Rev.\ Lett.}  {\bf 64}, 717 (1990).\\
  E.~Brezin and V.~A.~Kazakov,
  ``{\it Exactly Solvable Field Theories of Closed Strings,}''
  {\it Phys.\ Lett.\ B} {\bf 236}, 144 (1990).\\
  D.~Ghoshal,
  ``{\it p-adic string theories provide lattice discretization to the ordinary string worldsheet,}''
 {\it Phys.\ Rev.\ Lett.}  {\bf 97}, 151601 (2006).





  \bibitem{Biswas:2004qu}
  T.~Biswas, M.~Grisaru and W.~Siegel,
  ``{\it Linear Regge trajectories from worldsheet lattice parton field theory,}''
 {\it Nucl.\ Phys.\ B} {\bf 708}, 317 (2005)
  [hep-th/0409089].


\bibitem{Krasnov}
K.~Krasnov,
  ``{\it Renormalizable Non-Metric Quantum Gravity?,}''
  hep-th/0611182.\\
   K.~Krasnov,
 ``{\it Non-Metric Gravity I: Field Equations,}''
  {\it Class.\ Quant.\ Grav.}  {\bf 25}, 025001 (2008)
  [gr-qc/0703002].

\bibitem{Siegel:2003vt}
  W.~Siegel,
  ``{\it Stringy gravity at short distances,}''
  hep-th/0309093.



  \bibitem{Tseytlin:1995uq}
  A.~A.~Tseytlin,
  ``{\it On singularities of spherically symmetric backgrounds in string theory,}''
  {\it Phys.\ Lett.\ B} {\bf 363}, 223 (1995)
  [hep-th/9509050].


   \bibitem{Biswas:2005qr}
  T.~Biswas, A.~Mazumdar and W.~Siegel,
  ``{\it Bouncing universes in string-inspired gravity,}''
  {\it JCAP} {\bf 0603}, 009 (2006)
  [hep-th/0508194].
  
  


 \bibitem{Biswas:2010zk}
  T.~Biswas, T.~Koivisto and A.~Mazumdar,
  ``{\it Towards a resolution of the cosmological singularity in non-local higher derivative theories of gravity,}''
  {\it JCAP} {\bf 1011}, 008 (2010)
  [arXiv:1005.0590 [hep-th]].


  \bibitem{Biswas:2012bp}
  T.~Biswas, A.~S.~Koshelev, A.~Mazumdar and S.~Y.~Vernov,
  ``{\it Stable bounce and inflation in non-local higher derivative cosmology,}''
  {\it JCAP} {\bf 1208}, 024 (2012)
  [arXiv:1206.6374 [astro-ph.CO]].


  \bibitem{Craps:2014wga}
  B.~Craps, T.~De Jonckheere and A.~S.~Koshelev,
  ``{\it Cosmological perturbations in non-local higher-derivative gravity,}''
  arXiv:1407.4982 [hep-th].
  
\bibitem{Conroy:2014dja}
  A.~Conroy, A.~S.~Koshelev and A.~Mazumdar,
  ``{\it Geodesic completeness and homogeneity condition for cosmic inflation,}''
  {\it Phys.\ Rev.\ D} {\bf 90}, no. 12, 123525 (2014) [arXiv:1408.6205 [gr-qc]].
  
\bibitem{Biswas:2011ar}
  T.~Biswas, E.~Gerwick, T.~Koivisto and A.~Mazumdar,
  ``{\it Towards singularity and ghost free theories of gravity,}''
  {\it Phys.\ Rev.\ Lett.} {\bf 108}, 031101 (2012)
  [arXiv:1110.5249 [gr-qc]].

  \bibitem{Biswas:2013kla}
  T.~Biswas, T.~Koivisto and A.~Mazumdar,
  ``{\it Nonlocal theories of gravity: the flat space propagator,}''
  arXiv:1302.0532 [gr-qc].
  
  \bibitem{Modesto:2014eta}
  L.~Modesto, T.~d.~P.~Netto and I.~L.~Shapiro,
  ``{\it On Newtonian singularities in higher derivative gravity models,}''
  arXiv:1412.0740 [hep-th].
  
  \bibitem{Eliezer:1989cr} 
  D.~A.~Eliezer and R.~P.~Woodard,
  ``{\it The Problem of Nonlocality in String Theory,}''
  {\it Nucl.\ Phys.\ B} {\bf 325}, 389 (1989).
  

 
\bibitem{BT}  
 T.~Biswas and S.~Talaganis, ``{\it String-Inspired Infinite-Derivative Theories of Gravity: A Brief Overview,}'' {\it Mod. Phys. Lett. A}, Vol. {\bf 30}, Nos. 3 \& 4 (2015) 1540009 [arXiv:1412.4256 [gr-qc]].

\bibitem{Tomboulis}
  E.~Tomboulis,
  ``{\it Renormalizability and Asymptotic Freedom in Quantum Gravity,}''
 {\it Phys.\ Lett.\ B} {\bf 97}, 77 (1980).\\
  E.~T.~Tomboulis,
  ``{\it Renormalization And Asymptotic Freedom In Quantum Gravity,}''
  In *Christensen, S.m. ( Ed.): Quantum Theory Of Gravity*, 251-266 and Preprint - TOMBOULIS, E.T. (REC.MAR.83) 27p.\\
  E.~T.~Tomboulis,
  ``{\it Superrenormalizable gauge and gravitational theories,}''
  hep-th/9702146.
\bibitem{Modesto}
  L.~Modesto,
  ``{\it Super-renormalizable Quantum Gravity,}''
  {\it Phys.\ Rev.\ D} {\bf 86}, 044005 (2012)
  [arXiv:1107.2403 [hep-th]].\\
L.~Modesto and L.~Rachwal, ``{\it Super-renormalizable \& Finite Gravitational Theories,}'' arXiv:1407.8036 [hep-th].
\bibitem{BG}
  A.~O.~Barvinsky and Y.~.V.~Gusev,
  ``{\it New representation of the nonlocal ghost-free gravity theory,}''
  arXiv:1209.3062 [hep-th].\\
  A.~O.~Barvinsky,
  ``{\it Aspects of Nonlocality in Quantum Field Theory, Quantum Gravity and Cosmology,}''
  arXiv:1408.6112 [hep-th].

 \bibitem{Anslemi}
 D.~Anselmi,
  ``{\it Properties Of The Classical Action Of Quantum Gravity,}''
  {\it JHEP} {\bf 1305}, 028 (2013)
  [arXiv:1302.7100 [gr-qc]].
  \bibitem{Moffat-qg}
  J.~W.~Moffat,
  ``{\it Ultraviolet Complete Quantum Gravity,}''
  {\it Eur.\ Phys.\ J.\ Plus} {\bf 126}, 43 (2011)
  [arXiv:1008.2482 [gr-qc]].\\
L.~Modesto, J.~W.~Moffat and P.~Nicolini,
 ``{\it Black holes in an ultraviolet complete quantum gravity,}''
 {\it Phys.\ Lett.\ B} {\bf 695}, 397 (2011)
 [arXiv:1010.0680 [gr-qc]].


\bibitem{Efimov}
  G.~V.~Efimov,
  ``{\it Non-Local Quantum Theory of the Scalar Field,}''
  {\it Commun.\ Math.\ Phys.}  {\bf 5}, 42 (1967). \\
  V.~A.~Alebastrov and G.~V.~Efimov,
  ``{\it A proof of the unitarity of S matrix in a nonlocal quantum field theory,}''
  {\it Commun.\ Math.\ Phys.} {\bf 31}, 1 (1973). \\
V.~A.~Alebastrov and G.~V.~Efimov,
  ``{\it Causality In The Quantum Field Theory With The Nonlocal Interaction,}''
  {\it Commun.\ Math.\ Phys.}  {\bf 38}, 11 (1974).
D.~Anselmi,
  ``{\it Weighted scale invariant quantum field theories,}''
  {\it JHEP} {\bf 0802}, 051 (2008)
  [arXiv:0801.1216 [hep-th]].

\bibitem{Tomboulis:1983sw}
  E.~T.~Tomboulis,
  ``{\it Unitarity in Higher Derivative Quantum Gravity,}''
  {\it Phys.\ Rev.\ Lett.}  {\bf 52}, 1173 (1984).

\bibitem{Biswas:2014yia}
  T.~Biswas and N.~Okada,
  ``{\it Towards LHC Physics with Non-local Standard Model,}''
  arXiv:1407.3331 [hep-ph].


  \bibitem{Biswas:2009nx}
  T.~Biswas, J.~A.~R.~Cembranos and J.~I.~Kapusta,
  ``{\it Finite Temperature Solitons in Non-Local Field Theories from p-Adic Strings,}''
  {\it Phys.\ Rev.\ D} {\bf 82}, 085028 (2010)
  [arXiv:1006.4098 [hep-th]]. \\
T.~Biswas, J.~A.~R.~Cembranos and J.~I.~Kapusta,
  ``{\it Thermal Duality and Hagedorn Transition from p-adic Strings,}''
  {\it Phys.\ Rev.\ Lett.} {\bf 104}, 021601 (2010)
  [arXiv:0910.2274 [hep-th]]. \\
  T.~Biswas, J.~A.~R.~Cembranos and J.~I.~Kapusta,
  ``{\it Thermodynamics and Cosmological Constant of Non-Local Field Theories from p-Adic Strings,}''
  {\it JHEP} {\bf 1010}, 048 (2010)
  [arXiv:1005.0430 [hep-th]].


\bibitem{Moffat:1990jj}
  J.~W.~Moffat,
  ``{\it Finite nonlocal gauge field theory,}''
  {\it Phys.\ Rev.\ D} {\bf 41}, 1177 (1990). \\
D.~Evens, J.~W.~Moffat, G.~Kleppe and R.~P.~Woodard,
  ``{\it Nonlocal regularizations of gauge theories,}''
  {\it Phys.\ Rev.\ D} {\bf 43}, 499 (1991).

\bibitem{Bluhm}
R.~Bluhm,
  ``{\it Particle fields at finite temperature from string field theory,}''
  {\it Phys.\ Rev.\ D} {\bf 43}, 4042 (1991).


\bibitem{Reddy}
T.~Biswas, J.~Kapusta and A.~Reddy,
  ``{\it Thermodynamics of String Field Theory Motivated Nonlocal Models,}''
  {\it JHEP} {\bf 1212}, 008 (2012)
  [arXiv:1201.1580 [hep-th]].





\bibitem{Stelle:1976gc}
  K.~S.~Stelle,
  ``{\it Renormalization of Higher Derivative Quantum Gravity,}''
 {\it Phys.\ Rev.\ D} {\bf 16}, 953 (1977).

\bibitem{Goroff:1985th}
  M.~H.~Goroff and A.~Sagnotti,
  ``{\it The Ultraviolet Behavior of Einstein Gravity,}''
  {\it Nucl.\ Phys.\ B} {\bf 266}, 709 (1986).

  \bibitem{Goroff:1985sz}
  M.~H.~Goroff and A.~Sagnotti,
  ``{\it Quantum Gravity At Two Loops,}''
  {\it Phys.\ Lett.\ B} {\bf 160}, 81 (1985).






  \bibitem{Biswas:2013cha}
  T.~Biswas, A.~Conroy, A.~S.~Koshelev and A.~Mazumdar,
  ``{\it Generalized ghost-free quadratic curvature gravity,}''
  {\it Class.\ Quant.\ Grav.} {\bf 31}, 015022 (2014)
  [{\it Erratum-ibid.} {\bf 31}, 159501 (2014)]
  [arXiv:1308.2319 [hep-th]].\\
  H.~J.~Schmidt,
 ``{\it Variational Derivatives Of Arbitrarily High Order And Multiinflation
  Cosmological Models,}''
{\it Class.\ Quant.\ Grav.} {\bf 7} (1990) 1023.



   \bibitem{peter}
  P.~Van Nieuwenhuizen,
  ``{\it On ghost-free tensor lagrangians and linearized gravitation,}''
{\it Nucl.\ Phys.\ B} {\bf 60}, 478 (1973).
  


\end{thebibliography}
\end{document}